\documentclass[conference]{IEEEtran}
\IEEEoverridecommandlockouts
\usepackage{cite}
\usepackage{amsmath,amssymb,amsfonts}
\usepackage{graphicx}
\usepackage{textcomp}
\usepackage{xcolor}
\usepackage{bm}   
\usepackage{booktabs}
\usepackage{siunitx}
\usepackage{multirow}
\usepackage{algorithm}
\usepackage{algpseudocode}
\usepackage{colortbl}
\usepackage{makecell}
\usepackage{xeCJK}

\makeatletter
\newcommand{\linebreakand}{%
  \end{@IEEEauthorhalign}
  \hfill\mbox{}\par
  \mbox{}\hfill
  \begin{@IEEEauthorhalign}
}
\makeatother

\usepackage{listings}

\definecolor{promptbg}{HTML}{F8F9FA}
\definecolor{promptframe}{HTML}{DEE2E6}

\lstdefinestyle{promptstyle}{
    backgroundcolor=\color{promptbg},
    rulecolor=\color{promptframe},
    frame=single,
    framesep=5pt,
    basicstyle=\small\ttfamily,
    breaklines=true,       
    breakatwhitespace=true,
    columns=fullflexible,
    keepspaces=true,
    showstringspaces=false,
    tabsize=2,
    aboveskip=10pt,
    belowskip=10pt
}
\def\BibTeX{{\rm B\kern-.05em{\sc i\kern-.025em b}\kern-.08em
    T\kern-.1667em\lower.7ex\hbox{E}\kern-.125emX}}
\begin{document}

\title{SCAFFOLD-CEGIS: Preventing Latent Security Degradation in LLM-Driven Iterative Code Refinement}

\author{
\parbox{\textwidth}{\centering

\begin{minipage}[t]{0.47\textwidth}
\centering
{\fontsize{11}{13}\selectfont 1\textsuperscript{st} Yi Chen\par}
\vspace{2pt}
{\fontsize{11}{12}\selectfont
Chengdu Institute of Computer Application\par
Chinese Academy of Sciences\par
Chengdu, China\par
University of Chinese Academy of Sciences\par
Beijing, China\par
chenyi24@mail.ucas.ac.cn\par
}
\end{minipage}\hfill
\begin{minipage}[t]{0.47\textwidth}
\centering
{\fontsize{11}{13}\selectfont 2\textsuperscript{nd} Yun Bian\par}
\vspace{2pt}
{\fontsize{11}{12}\selectfont
Chengdu Institute of Computer Application\par
Chinese Academy of Sciences\par
Chengdu, China\par
University of Chinese Academy of Sciences\par
Beijing, China\par
bianyun@casit.com.cn\par
}
\end{minipage}

\vspace{0.8em}

\begin{minipage}[t]{0.31\textwidth}
\centering
{\fontsize{10}{12}\selectfont 3\textsuperscript{rd} Haiquan Wang\par}
\vspace{2pt}
{\fontsize{9.5}{10.5}\selectfont
Chengdu Institute of Computer\par
Application\par
Chinese Academy of Sciences\par
Chengdu, China\par
University of Chinese Academy of Sciences\par
Beijing, China\par
wanghaiquan22@mails.ucas.ac.cn\par
}
\end{minipage}\hfill
\begin{minipage}[t]{0.31\textwidth}
\centering
{\fontsize{10}{12}\selectfont 4\textsuperscript{th} Shihao Li\par}
\vspace{2pt}
{\fontsize{9.5}{10.5}\selectfont
Chengdu Institute of Computer\par
Application\par
Chinese Academy of Sciences\par
Chengdu, China\par
University of Chinese Academy of Sciences\par
Beijing, China\par
lishihao25@mails.ucas.ac.cn\par
}
\end{minipage}\hfill
\begin{minipage}[t]{0.31\textwidth}
\centering
{\fontsize{10}{12}\selectfont 5\textsuperscript{th} Zhe Cui$^*$\par}
\vspace{2pt}
{\fontsize{9.5}{10.5}\selectfont
Chengdu Institute of Computer\par
Application\par
Chinese Academy of Sciences\par
Chengdu, China\par
University of Chinese Academy of Sciences\par
Beijing, China\par
cuizhe@casit.com.cn\par
}
\end{minipage}

}
}

\maketitle

\begin{abstract}
The application of large language models to code generation has evolved from one-shot generation to iterative refinement, yet the evolution of security throughout iteration remains insufficiently understood. Through comparative experiments on three mainstream LLMs, this paper reveals the iterative refinement paradox: specification drift during multi-objective optimization causes security to degrade gradually over successive iterations. Taking GPT-4o as an example, 43.7\% of iteration chains contain more vulnerabilities than the baseline after ten rounds, and cross-model experiments show that this phenomenon is prevalent.

Further analysis shows that simply introducing static application security testing (SAST) gating cannot effectively suppress degradation; instead, it increases the latent security degradation rate from 12.5\% under the unprotected baseline to 20.8\%. The root cause is that static-analysis rules cannot cover structural degradations such as the removal of defensive logic or the weakening of exception handling. To address this problem, we propose the SCAFFOLD-CEGIS framework. Drawing on the counterexample-guided inductive synthesis (CEGIS) paradigm, the framework adopts a multi-agent collaborative architecture that transforms security constraints from implicit prompts into explicit verifiable constraints. It automatically identifies and solidifies security-critical elements as hard constraints through semantic anchoring, enforces safety monotonicity through four-layer gated verification, and continuously assimilates experience from failures. Comparative experiments against six existing defense methods show that the full framework reduces the latent security degradation rate to 2.1\% and achieves a safety monotonicity rate of 100\%. These results indicate that LLM-driven iterative code refinement entails an inherent risk of security degradation, that static-analysis gating may further induce a pseudo-security effect, and that explicit semantic constraints are necessary to suppress latent security degradation during iterative refinement.
\end{abstract}


\section{Introduction}

The use of large language models in software engineering has expanded from one-shot code generation to iterative code refinement. Tools such as GitHub Copilot and Cursor allow developers to interact with LLMs over multiple rounds to improve functionality, optimize performance, and fix defects. However, prior work has focused primarily on single-shot generation quality, while the evolution of security throughout iterative refinement remains insufficiently studied.

\subsection{The Iterative Refinement Paradox}

Current AI-assisted programming workflows implicitly assume that multi-turn interaction progressively improves code quality. Our empirical results show that, in the absence of strong external constraints, iterative LLM refinement can instead degrade code security.

The root cause is specification drift in multi-objective optimization. In each iteration, the LLM jointly optimizes competing objectives, including functional completeness, performance efficiency, maintainability, and security. When security constraints exist only as soft prompts, the optimization trajectory gradually departs from the intended security specification, for example by removing validation logic, breaking API compatibility, or introducing resource leaks.

Experiments on 96 iteration chains with GPT-4o support this claim. When all vulnerability severities are counted ($DR_{all}$), a substantial share of chains already exceed the baseline vulnerability count by iteration 10. Even under explicit security-hardening prompts, the degradation rate remains 28.6\%. Detailed settings and vulnerability trajectories are presented in Section~\ref{sec:exp_setup}.

\subsection{Coverage Blind Spots of SAST Gating}

A natural mitigation strategy is to run static analysis after each iteration and reject changes that introduce new vulnerabilities. However, static analyzers have coverage blind spots. We define one class of degradation as \emph{latent security degradation}: security-critical elements are removed or weakened during iteration without triggering SAST rules. Typical cases include deleting validation functions without introducing new SAST-detectable patterns, weakening exception handling logic, or bypassing permission checks along paths not covered by the analyzer.

SAST detection logic is grounded in known vulnerability patterns and cannot directly capture accidental removal or weakening of security-critical elements. Once an LLM removes validation or sanitization logic during refactoring, degradation remains undetected if the remaining code does not match rule-based vulnerability signatures.

\subsection{Contributions}

To address latent security degradation, this paper introduces SCAFFOLD-CEGIS, a multi-agent framework that adopts the structural principle of CEGIS (Counterexample-Guided Inductive Synthesis). The framework protects security-critical elements via semantic anchoring and closes key SAST coverage gaps.

The main contributions are as follows.

\textbf{Iterative refinement paradox and SAST blind spots.} Controlled experiments on 96 GPT-4o iteration chains reveal systematic security degradation in iterative LLM refinement, and show that static analyzers miss structural degradations such as deleted defensive logic and weakened exception handling.

\textbf{SCAFFOLD-CEGIS framework.} We propose a CEGIS-inspired multi-agent architecture that transforms security constraints from implicit prompt instructions into explicit, verifiable hard constraints. The framework combines semantic anchoring, four-layer gated verification, and failure assimilation. Semantic anchoring mines security-critical functions, defense patterns, and API-compatibility constraints and solidifies them as invariants. Four-layer gating checks correctness, safety monotonicity, change budget, and anchor integrity for each candidate. Failure assimilation extracts structured lessons from rejected candidates to guide subsequent generation.

\textbf{Cross-language and cross-model validation.} Experiments across multiple mainstream LLMs and languages confirm the reproducibility of the iterative refinement paradox. Ablation studies isolate the contribution of each component. The full framework reduces latent security degradation from 20.8\% to 2.1\% and attains 100\% safety monotonicity under our setup, while preserving code evolution capability.

The remainder of this paper is organized as follows. Section II reviews related work. Section III presents the SCAFFOLD-CEGIS framework. Section IV reports the experimental setup and results. Section V discusses limitations and future work. Section VI concludes.

\section{Related Work}
\label{sec:related_work}

\subsection{Security Evaluation of LLM Code Generation}
As large language models (LLMs) become deeply integrated into practical development workflows, including code completion, natural-language-to-code generation, and automated bug fixing (e.g., GitHub Copilot and Amazon CodeWhisperer), the security properties of model-generated code have attracted sustained academic scrutiny: while satisfying functional requirements, do these models reliably adhere to secure coding practices?


Existing empirical studies analyze this problem mainly from two dimensions: vulnerability-introduction tendency and prompt robustness. Multiple independent studies show that, without explicit security constraints, mainstream LLMs frequently reproduce vulnerable coding patterns present in training data. Perry et al.~\cite{perry2023users} report through large-scale user studies that developers using coding assistants are more likely to accept insecure suggestions and often fail to identify the defects. Khoury et al.~\cite{khoury2023secure} show systematic weaknesses in key management and randomness generation in cryptographic tasks. Siddiq et al.~\cite{siddiq2024sallm} further quantify this risk with the SALLM benchmark, using secure@k (the probability that all $k$ samples are vulnerability-free) and vulnerable@k (the probability that at least one sample is vulnerable) in SQL injection and XSS defense scenarios, and find frequent omission of critical input-validation logic (e.g., parameterized queries or output escaping) even when models are explicitly instructed to generate secure code. Pearce et al.~\cite{pearce2025asleep} expose a deeper issue: models may not only generate vulnerable code but also provide incorrect security explanations, claiming insecure implementations are safe.

To transform these qualitative observations into measurable evidence, recent work has focused on standardized benchmarks. Hajipour et al.~\cite{hajipour2024codelmsec} propose \texttt{CodeLM-Sec}, which stress-tests black-box models with adversarial prompt templates targeting specific vulnerability classes (e.g., buffer overflow and command injection). Li et al.~\cite{li2024attribution} focus on attribution analysis, aiming to separate vulnerabilities caused by pretraining-data contamination from those caused by alignment failure at inference. Tihanyi et al.~\cite{tihanyi2025secure} and \texttt{SecCodeBench}~\cite{seccodebench2025} establish cross-language and cross-model comparative benchmarks and report vulnerability-generation rates across fixed task sets. Collectively, these efforts mark a shift from case-based analysis to benchmark-driven evaluation in LLM code security.

For mitigation, current studies follow two major directions: inference-time alignment and data/training alignment. The former filters unsafe outputs using security-oriented prompting, constrained decoding, or static-analysis feedback loops. He et al.~\cite{he2025cosefa} use static-analysis diagnostics as feedback signals for model self-correction, while Tony et al.~\cite{tony2025prompting} investigate chain-of-thought prompting for security reasoning. The latter suppresses insecure-generation patterns at the distribution level through high-quality secure-code datasets or instruction tuning. He et al.~\cite{he2024instruction} release an instruction dataset for security repair, and De et al.~\cite{de2024enhanced} study the effectiveness of RLHF in optimizing security preferences.

Most of the above methods target the security of single-shot generation. A recent survey~\cite{bouzid2025assessing} notes that existing strategies focus on a static ``generate-detect'' loop and insufficiently account for the dynamic evolution of security properties in multi-turn interaction. In particular, when developers iteratively optimize functionality or performance with LLMs, early security constraints are easily diluted or forgotten in subsequent context windows, yielding the paradox of ``functional improvement with security degradation.''

\subsection{Iterative Program Synthesis and Feedback Mechanisms}
Iterative program synthesis seeks target programs through a closed loop of generation, verification, feedback, and regeneration. Classical paradigms such as Counterexample-Guided Inductive Synthesis (CEGIS)~\cite{solar2006combinatorial,solar2008program} rely on formal verifiers to provide precise counterexamples that contract the search space. Building on this idea, Syntax-Guided Synthesis (SyGuS)~\cite{alur2013syntax} introduces grammar constraints to further bound candidate programs and keep synthesis within decidable regimes. A defining characteristic of these traditional methods is \emph{strong external constraints}: the iteration process is tightly specified, and any candidate that violates the specification is immediately eliminated.

In LLM-driven synthesis, iterative mechanisms are typically instantiated as self-debugging or multi-agent collaboration. The \texttt{Reflexion} framework proposed by Shinn et al.~\cite{shinn2023reflexion} improves subsequent attempts by prompting models to reflect on natural-language trajectories of failed execution and by accumulating failure experience in memory. Similarly, reinforcement-learning-based approaches~\cite{le2022coderl} aim to improve model utilization of execution feedback such as compilation errors and unit-test failures. Recent studies~\cite{madaan2023self} further distinguish between post-execution repair and in-execution repair, and identify failure modes caused by incomplete test coverage, noisy feedback, and ambiguous error localization.

In existing LLM iterative synthesis studies, feedback is sourced primarily from \emph{functional correctness}, i.e., whether code compiles or passes unit tests. Security properties (e.g., memory safety, access control, and completeness of input validation) are difficult to cover with unit tests alone, and usually lack CEGIS-style formal specifications as hard constraints. Consequently, under weak security feedback, models tend to overfit local functional objectives during iteration, unintentionally removing defensive logic or introducing new attack surfaces during refactoring. This security-objective drift induced by weak feedback is the core cause of the iterative refinement paradox defined in this paper.

\section{SCAFFOLD-CEGIS Framework}

\subsection{Theoretical Foundations}

We model iterative code refinement as a state-transition system. Let $\mathcal{P}$ denote the set of programs and $\mathcal{T}$ the set of tasks. The transition function $\delta: \mathcal{P} \times \mathcal{T} \rightarrow \mathcal{P}$ maps program $P_i$ to $P_{i+1}$ given task $\tau$. An iteration chain is defined as the state sequence $\mathcal{C} = \langle P_0, P_1, \ldots, P_n \rangle$.

The security metric $V: \mathcal{P} \rightarrow \mathbb{N}$ maps a program to its vulnerability count. Safety monotonicity over an iteration chain $\mathcal{C}$ is defined in Eq.~(\ref{eq:safety_monotonicity_def}).
\begin{equation}
\forall i,\; V(P_{i+1}) \leq V(P_i)
\label{eq:safety_monotonicity_def}
\end{equation}
The degradation rate $DR$ and safety monotonicity rate $SMR$ are defined in Eq.~(\ref{eq:dr_def}) and Eq.~(\ref{eq:smr_def}), respectively.
\begin{equation}
DR = \frac{|\{\mathcal{C} : V(P_n) > V(P_0)\}|}{|\mathcal{C}_{all}|}
\label{eq:dr_def}
\end{equation}
\begin{equation}
SMR = \frac{\sum_{\mathcal{C}} |\{i : V(P_{i+1}) \leq V(P_i)\}|}{\sum_{\mathcal{C}} n_{\mathcal{C}}}
\label{eq:smr_def}
\end{equation}

SCAFFOLD-CEGIS constrains state transitions through a gating function. The security specification $\Phi$ is an explicit constraint set generated by semantic anchoring, including security-critical invariants, API signatures, and defense patterns. The gate function $G$ is formally defined in Eq.~(\ref{eq:gate_signature}), and internally implemented as the conjunction of four sub-checks (Eq.~(\ref{eq:gate_decomposition})): correctness, safety monotonicity, diff budget, and anchor integrity.
\begin{equation}
G: \mathcal{P} \times \mathcal{P} \times \Phi \rightarrow \{\mathrm{accept}, \mathrm{reject}\}
\label{eq:gate_signature}
\end{equation}
\begin{equation}
\begin{aligned}
G(P, P', \Phi) ={}& V_{corr}(P') \land V_{sec}(P, P') \\
&\land V_{diff}(P, P') \land V_{anchor}(P', \Phi)
\end{aligned}
\label{eq:gate_decomposition}
\end{equation}
When $\delta_{max} = 0$, the gated iteration chain satisfies Eq.~(\ref{eq:monotonicity_guarantee}).
\begin{equation}
\delta_{max} = 0 \Rightarrow \forall i,\; V(P_{i+1}) \leq V(P_i)
\label{eq:monotonicity_guarantee}
\end{equation}

\subsection{Framework Overview}

\begin{figure*}[t]
\centering
\includegraphics[width=\textwidth]{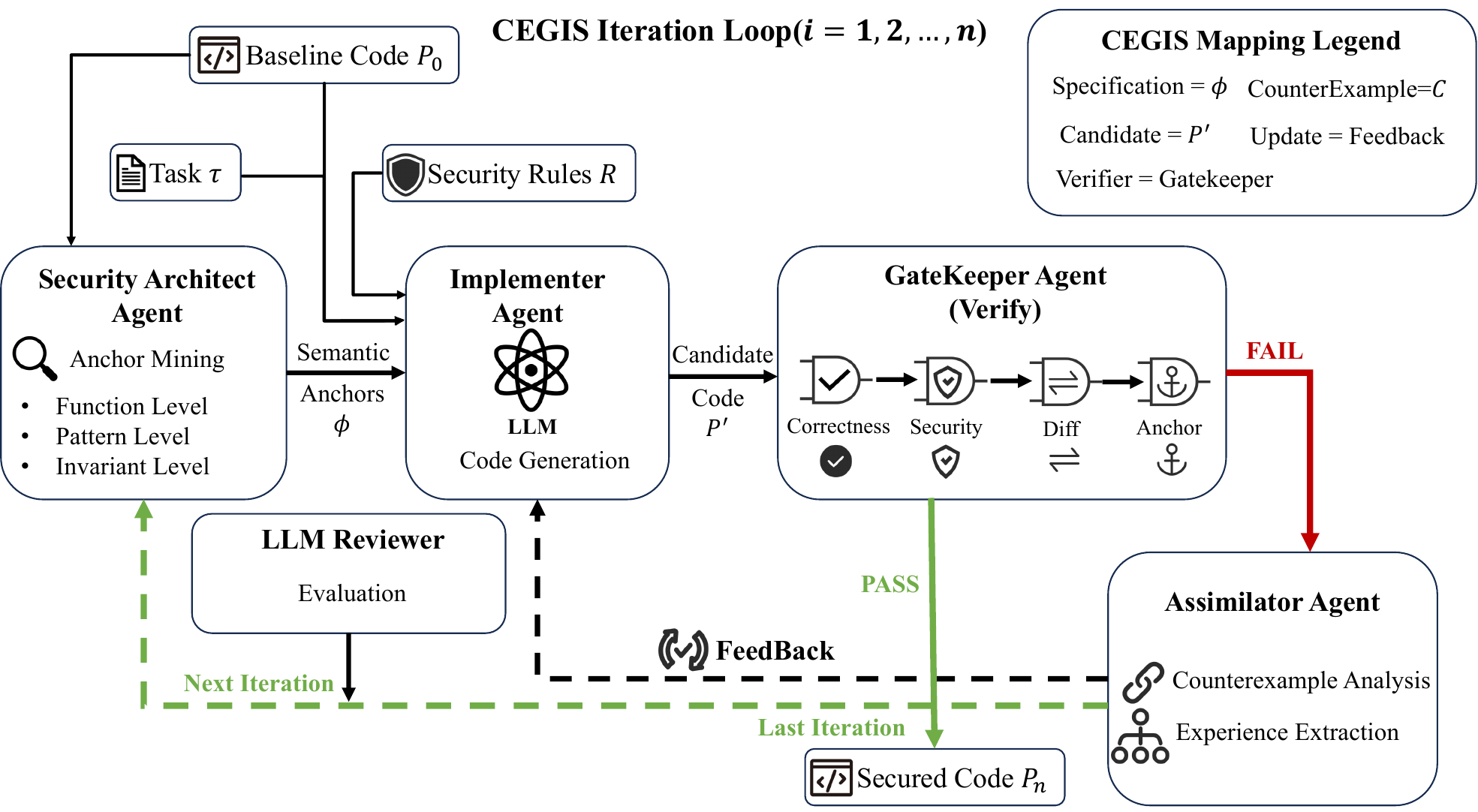}
\caption{Overview of the SCAFFOLD-CEGIS architecture. The system follows a CEGIS-style workflow with four collaborative agents: (1) SecurityArchitectAgent mines semantic anchors from code to construct the security specification $\Phi$; (2) ImplementerAgent generates candidate code $P'$ under anchor constraints; (3) GatekeeperAgent validates candidates with four-layer gating; (4) AssimilatorAgent extracts reusable feedback from failures.}
\label{fig:overview}
\end{figure*}

SCAFFOLD-CEGIS adopts a multi-agent collaborative architecture that converts security constraints from implicit prompts into explicit verifiable constraints. The framework includes four agents: SecurityArchitectAgent mines security-critical elements from code and constructs the semantic anchor set $\Phi$; ImplementerAgent generates candidate code $P'$ under anchor constraints; GatekeeperAgent performs four-layer gating to validate correctness and safety monotonicity of each change; AssimilatorAgent analyzes rejected attempts and extracts reusable experience for subsequent iterations. Figure~\ref{fig:overview} illustrates the overall architecture.

This framework borrows the iterative structure of CEGIS, i.e., a closed loop of candidate generation, verification, feedback, and regeneration, while relaxing the formal counterexample requirement. In classical CEGIS, verifiers are based on formal specifications and SMT solving, and the resulting counterexamples precisely shrink the candidate search space. In our framework, the verifier is replaced by a combination of static analysis and semantic-anchor checking, and counterexamples are represented as structured failure reports produced after gate rejection, including failure type, violated constraints, and code locations, and passed to ImplementerAgent in natural language. This substitution trades formal completeness for practical applicability in real software-engineering settings, enabling handling of security properties that are difficult to fully formalize.

The execution flow follows this iterative process. SecurityArchitectAgent analyzes the current code and produces the semantic anchor set. ImplementerAgent then generates a candidate satisfying the current refinement task under hard anchor constraints. GatekeeperAgent performs four-layer verification of correctness and safety monotonicity. Upon verification failure, the system decides between retry and rollback according to failure type, while AssimilatorAgent records failure causes and extracts structured lessons. The security specification $\Phi = \langle \mathcal{I}, \mathcal{A}, \mathcal{R} \rangle$ consists of invariant anchors $\mathcal{I}$, API-signature anchors $\mathcal{A}$, and a security rule base $\mathcal{R}$. Anchors are updated dynamically as code evolves, and stale constraints are removed by a TTL mechanism.

Algorithm~\ref{alg:scaffold} summarizes the core loop. In each iteration, SecurityArchitectAgent mines semantic anchors, ImplementerAgent generates a candidate, and GatekeeperAgent executes four-layer verification. Accepted candidates are committed; rejected candidates trigger retry or rollback according to failure type. AssimilatorAgent extracts experience from failures and feeds it into subsequent iterations. After all iterations, an LLM Reviewer performs semantic review on the final code; its output is used only for SSDR and RDR evaluation and does not participate in gate decisions.
The algorithm guarantees $V(P_n) \leq V(P_0)$ when $\delta_{max} = 0$. In our experiments, we set $\delta_{max} = 0$, requiring each iteration to introduce no new SAST-detectable vulnerabilities.

\begin{algorithm}[t]
\caption{SCAFFOLD-CEGIS Iterative Procedure}
\label{alg:scaffold}
\begin{algorithmic}[1]
\Require Baseline code $P_0$, refinement task sequence $\mathcal{T} = \{\tau_1, \ldots, \tau_n\}$, maximum retries $R_{max}$, security tolerance threshold $\delta_{max}$
\Ensure Final code $P_n$ satisfies $V(P_n) \leq V(P_0)$
\State $P \gets P_0$
\State $\mathcal{K} \gets \emptyset$ \Comment{Failure knowledge base}
\State $S_0 \gets \textsc{LLMReview}(P_0)$ \Comment{Baseline latent-security issue count}
\For{$i = 1$ to $n$}
    \State $\Phi \gets \textsc{MineAnchors}(P)$ \Comment{SecurityArchitectAgent}
    \State $r \gets 0$
    \While{$r < R_{max}$}
        \State $P' \gets \textsc{Implement}(P, \tau_i, \Phi, \mathcal{K})$ \Comment{ImplementerAgent}
        \State $T_{result} \gets \textsc{RunTests}(P')$ \Comment{Correctness check}
        \State $\Delta V_{step} \gets V_{SAST}(P') - V_{SAST}(P)$ \Comment{Safety-monotonicity check}
        \State $\Delta L \gets |P'| - |P|$ \Comment{Diff-budget check}
        \State $A_{result} \gets \textsc{VerifyAnchors}(P', \Phi)$ \Comment{Anchor-integrity check}
        \If{$T_{result} = \texttt{PASS} \land \Delta V_{step} \leq \delta_{max} \land |\Delta L| \leq B \land A_{result} = \texttt{PASS}$}
            \State $P \gets P'$ \Comment{Accept change}
            \State \textbf{break}
        \Else
            \State $\mathcal{C} \gets \textsc{GenCE}(P, P', T_{result}, \Delta V_{step}, A_{result})$
            \State $\mathcal{K} \gets \mathcal{K} \cup \textsc{Assimilate}(\mathcal{C})$ \Comment{AssimilatorAgent}
            \State $r \gets r + 1$
        \EndIf
    \EndWhile
    \If{$r = R_{max}$}
        \State \textbf{continue} \Comment{Rollback; keep $P$ unchanged}
    \EndIf
\EndFor
\State $S_n \gets \textsc{LLMReview}(P)$ \Comment{Final latent-security issue count, for SSDR}
\State \Return $P$
\end{algorithmic}
\end{algorithm}

\subsection{Semantic Anchor Mining}

Semantic anchors are formal representations of security-critical elements in code, used to prevent accidental modification or deletion during iteration. SecurityArchitectAgent mines anchors from three levels.

Function-level anchors use regex matching to identify security-critical functions, such as validation functions \texttt{validate\_*}, sanitization functions \texttt{sanitize\_*}, and authentication functions \texttt{auth*}. AST parsing then extracts function signatures, including function names, parameter lists, and return types. For public APIs, required parameter lists are additionally recorded to prevent backward-compatibility breaks.

Relying only on naming conventions can miss security-critical elements; therefore, the system supplements naming-based detection with data-flow analysis. It maintains a list of sensitive sinks (e.g., SQL execution \texttt{execute}, file operations \texttt{open}, command execution \texttt{exec}, and deserialization \texttt{pickle.loads}). For each sink, the system traces the call chain upstream to identify guard functions between user input and the sink. Specifically, a function $f$ is marked as an anchor candidate if it lies on control-flow paths between input boundary and sensitive sink, contains conditional branching or exception throwing, and has data dependency between its parameters and sink parameters. This method captures non-standard defensive functions such as \texttt{isSafePath} and \texttt{assertAllowed}.

Pattern-level anchors identify defensive logic through code-pattern matching. The maintained security pattern library includes input-validation patterns such as \texttt{raise ValueError}, permission-check patterns such as \texttt{raise PermissionError}, and parameterized-query patterns such as \texttt{execute(..., ?)}. After a successful match, the system extracts the corresponding snippet and context to generate substring or regex anchors.

Invariant-level anchors are derived from a security-invariant library. For each task category, the system maintains a set of invariants; for example, database tasks require all SQL queries to be parameterized, and file-processing tasks require all path operations to be normalized. These invariants are validated by static analysis after each iteration.

Each anchor has six attributes: \texttt{anchor\_type} (substring/regex/AST/function signature/invariant), \texttt{selector} (matching content), \texttt{lock\_level} (\texttt{hard}: reject on violation; \texttt{soft}: warning only), \texttt{priority} (four levels from \texttt{CRITICAL} to \texttt{LOW}), \texttt{created\_iteration} (creation time), and \texttt{confidence} (confidence score).

Anchor lifecycle is managed by TTL and migration protocols. \texttt{CRITICAL} and \texttt{HIGH} anchors do not expire by TTL. \texttt{MEDIUM} and \texttt{LOW} anchors are initialized with TTL values of 10 and 5 iterations, respectively; TTL is reset when the code element remains present, and migration is triggered when code is refactored but equivalent functionality remains.

When ImplementerAgent needs to refactor protected code, it may submit an anchor-migration request. GatekeeperAgent verifies whether the new anchor satisfies the same security invariants, covers the original defense scope, and preserves interface compatibility. If verification passes, the anchor is migrated to the new location with unchanged priority and constraint strength. Migration of \texttt{CRITICAL} anchors requires equivalence evidence or additional security review.

\subsection{Gated Verification Mechanism}

GatekeeperAgent implements four-layer gated verification. Each layer executes independently and returns \texttt{PASS}, \texttt{FAIL}, \texttt{FAIL\_RETRY}, or \texttt{WARN}. Checks are executed in priority order, and any critical failure terminates subsequent checks.

The correctness layer executes the test suite. A candidate is admitted only when all tests pass; otherwise, \texttt{FAIL\_RETRY} is returned.

The safety-monotonicity layer compares static-analysis results between consecutive iterations. Its quantitative definition is given in Eq.~(\ref{eq:gate_deltas}).
\begin{equation}
\begin{aligned}
\Delta_{CH} &= (C_{curr} + H_{curr}) - (C_{prev} + H_{prev}) \\
\Delta_{\rho} &= \rho_{curr} - \rho_{prev},\quad
\rho = \frac{1000 \times \text{SAST\_Risk}}{\max(\text{LOC}, 1)}
\end{aligned}
\label{eq:gate_deltas}
\end{equation}
The pass condition is defined in Eq.~(\ref{eq:gate_pass}).
\begin{equation}
\Delta_{CH} \leq \delta_{max} \land \Delta_{\rho} \leq \epsilon
\label{eq:gate_pass}
\end{equation}

The diff-budget layer limits per-iteration change scale, with constraints in Eq.~(\ref{eq:diff_budget_constraints}), to reduce risks from large refactors.
\begin{equation}
L_{add} \leq B_{add},\qquad L_{del} \leq B_{del}
\label{eq:diff_budget_constraints}
\end{equation}

The anchor-integrity layer verifies all hard-level anchors through substring search, regex matching, AST parsing, or semantic validation according to anchor type. Missing \texttt{CRITICAL} anchors trigger immediate rejection, missing \texttt{HIGH} anchors trigger retry, and lower-priority losses trigger warnings only.

Gate decision logic is as follows: accept the candidate when all checks pass or only warnings are present; trigger re-implementation when \texttt{FAIL\_RETRY} occurs and retry budget remains; rollback to the previous safe version when retries are exhausted or a \texttt{FAIL} is returned.

\subsection{Failure Assimilation and Feedback Learning}

AssimilatorAgent extracts reusable knowledge from rejected attempts. Failures are categorized by gate type: correctness failures capture test cases and error messages; safety-monotonicity failures record vulnerability types and locations; anchor-violation failures identify broken anchors; diff-budget failures record change-scale statistics.

Experience extraction relies on pattern recognition. When similar failures recur, the system produces generalized remediation guidance; for example, repeated anchor violations caused by deleting validation functions generate the rule ``avoid deleting functions whose names contain validate,'' and repeated safety-monotonicity failures caused by SQL string concatenation generate the rule ``use parameterized queries instead of string concatenation.'' These rules are stored in natural language and delivered as developer messages to ImplementerAgent in later iterations.

Feedback has two forms. Current-iteration failures generate immediate feedback with failure cause and fix direction. Historical failures are accumulated in a knowledge base, and similar cases are retrieved at the beginning of new iterations.

An anchor-feedback mechanism dynamically adjusts the anchor set. If an anchor is repeatedly violated across iterations but the change is eventually accepted, the system lowers its priority or downgrades it from hard to soft. If a code element appears frequently in failure cases, the system creates a new anchor for that element and raises its priority. This mechanism allows anchor sets to co-evolve with code.

\section{Experiments}

\subsection{Experimental Setup}
\label{sec:exp_setup}

The experiments are organized around three research questions. RQ1 examines whether the iterative refinement paradox holds across languages, task types, and models. RQ2 evaluates whether SCAFFOLD-CEGIS can convert empirically observed safety monotonicity into verifiable engineering guarantees. RQ3 isolates the independent contributions of semantic anchoring, gated verification, and failure assimilation through ablation.

The study follows a staged design. The observational study in the introduction fixes the model to GPT-4o and metric scope to $DR_{all}$, aiming to establish the existence of degradation and characterize its trajectory over iterations. RQ1 expands model diversity to three mainstream LLMs while keeping prompt templates and temperature settings consistent, to test reproducibility across models and languages. RQ2 and RQ3 continue under a unified protocol for method comparison and mechanism attribution.

The dataset contains 24 programming-task samples covering six development scenarios (database operations, input handling, authentication, resource management, cryptography, and path handling), with two samples per category in both Python and Java. Each iteration chain runs for 10 refinement rounds. Iteration strategies include feature enhancement, performance optimization, security hardening, and ambiguous requirements. Each sample is executed under 3 models, 2 languages, and 4 strategies, yielding 288 chains and 2,880 iteration steps in total.

The dataset is designed for diversity of security scenarios. Each of the six categories corresponds to a representative vulnerability context, covering SQL injection defense, input-validation boundaries, session/token validation, resource limits and concurrency control, secure randomness and cryptographic configuration, and path-traversal defense (Table~\ref{tab:dataset}). The two samples in each category differ in code complexity to cover both simple implementations and complex implementations with multi-layer defensive logic. All samples are self-contained single-file modules with complete defense logic and test suites, and their coding patterns are derived from analogous security components in mainstream open-source projects. Table~\ref{tab:dataset} reports scale and complexity statistics: lines of code (LOC, excluding comments and blanks) range from 25 to 392, cyclomatic complexity (CC) ranges from 4 to 145, and each sample includes 5 to 20 test cases. This scale diversity reduces dependence on a specific code size. The applicability of conclusions is limited to the covered task types; extension to multi-file projects or larger codebases requires further validation.

\begin{table}[t]
\centering
\caption{Dataset Statistics (4 samples per category: 2 languages $\times$ 2 samples)}
\label{tab:dataset}
\resizebox{\columnwidth}{!}{%
\begin{tabular}{llcc}
\toprule
Category & Security Focus & LOC Range & CC Range \\
\midrule
Database Operations & SQL Injection Defense & 61--191  & 15--73  \\
Input Handling & Input-Validation Boundaries & 90--392  & 41--145 \\
Authentication & Session/Token Validation & 25--287  & 4--98   \\
Resource Management & Resource Limits/Concurrency & 107--210 & 20--61  \\
Cryptography & Secure Randomness/Encryption & 30--60   & 5--7    \\
Path Handling & Path Traversal Defense & 77--177  & 20--40  \\
\midrule
\textbf{Overall} & & 25--392 (mean 124) & 4--145 (mean 37) \\
\bottomrule
\end{tabular}%
}
\end{table}

\begin{figure*}[t]
\centering
\begin{minipage}[b]{0.48\textwidth}
    \centering
    \includegraphics[width=\textwidth]{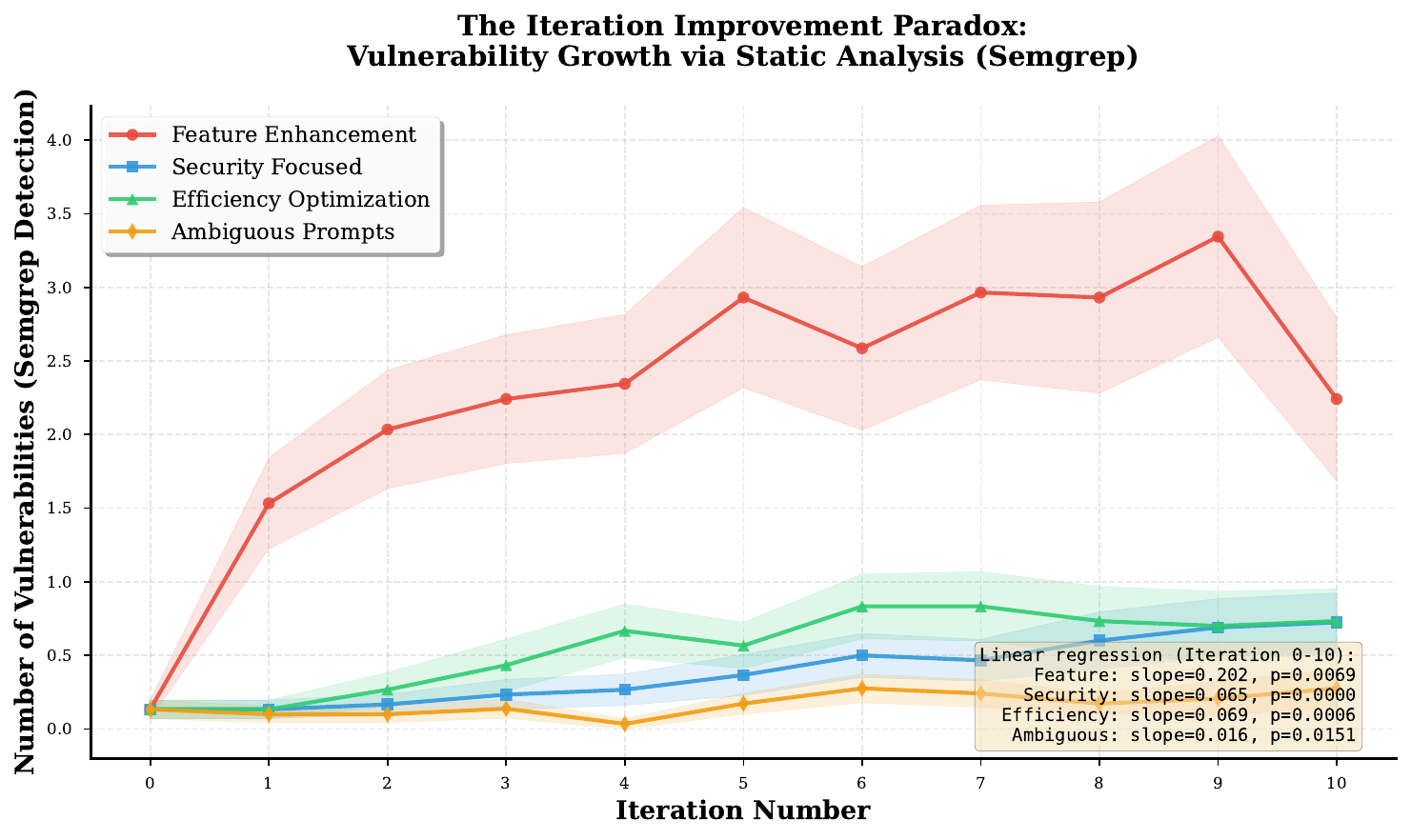}
    \vspace{-2mm}
    \centerline{(a) Semgrep-based static analysis}
\end{minipage}
\hfill
\begin{minipage}[b]{0.48\textwidth}
    \centering
    \includegraphics[width=\textwidth]{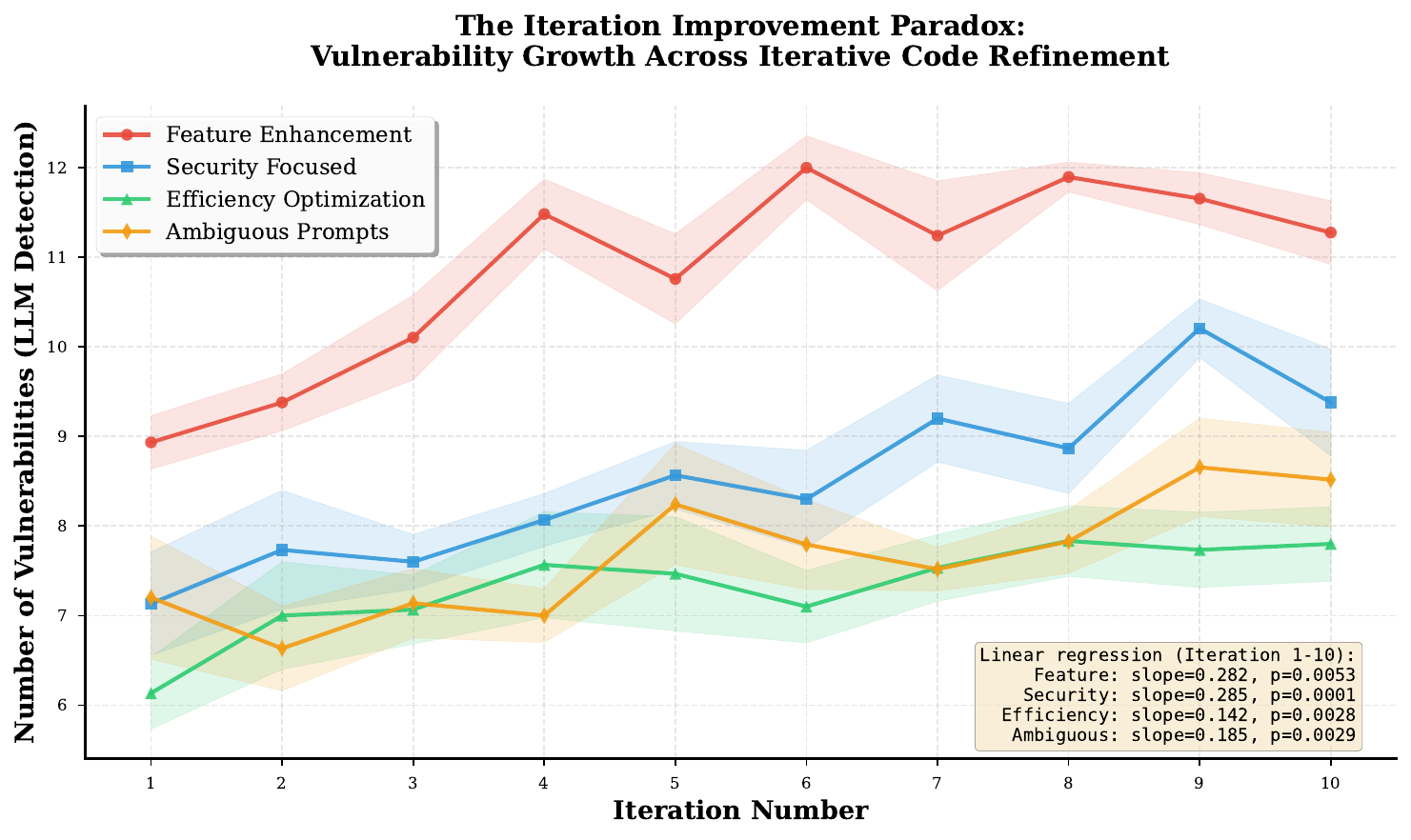}
    \vspace{-2mm}
    \centerline{(b) with LLM semantic review}
\end{minipage}
\vspace{-2mm}
\caption{Vulnerability trajectories under the iterative refinement paradox. (a) Growth in vulnerabilities detected by static analysis, where feature-enhancement prompts degrade most severely; (b) issues detected by LLM semantic review, with counts substantially higher than static-analysis findings.}
\label{fig:paradox}
\end{figure*}

The iterative refinement paradox reported in the introduction is based on an observational GPT-4o study using Semgrep with the \texttt{p/default} rule set. At iteration 10, 43.7\% of chains exceed baseline vulnerability counts. This result is used to establish the research problem, not for cross-model ranking. Table~\ref{tab:code_churn} reports degradation differences across strategies; correlations between code evolution volume (code churn) and degradation are consistently low across strategies. Figure~\ref{fig:paradox} shows vulnerability trajectories over iterations: feature-enhancement prompts exhibit the strongest degradation, and LLM semantic review detects substantially more issues than static analysis.

\begin{table}[t]
\centering
\caption{Code Evolution Volume (Code Churn) and Degradation Rate ($DR_{all}$) Across Strategies}
\label{tab:code_churn}
\resizebox{\columnwidth}{!}{%
\begin{tabular}{lcccc}
\toprule
Strategy & Avg. Churn (LOC) & Degradation Rate & Correlation Coefficient \\
\midrule
Feature Enhancement & 79.6 & 52.8\% & 0.156 \\
Performance Optimization & 40.6 & 38.2\% & 0.089 \\
Security Hardening & 36.4 & 28.6\% & 0.072 \\
Ambiguous Requirement & 60.5 & 44.5\% & 0.118 \\
\bottomrule
\end{tabular}
}
\end{table}

Table~\ref{tab:model_config} lists model versions used in experiments. All models share the same prompt template and temperature setting $T=0.7$, while other parameters remain default. To detect latent security defects beyond static-analysis coverage, we use Claude Opus 4.5 as an independent LLM reviewer. This reviewer is used only for offline evaluation and does not participate in any gating decisions, approximating manual code-review practice.

\begin{table}[t]
\centering
\caption{Model Version Configuration Used in Experiments}
\label{tab:model_config}
\begin{tabular}{lll}
\toprule
Alias & API Model Identifier & Provider \\
\midrule
GPT-4o & \texttt{gpt-4o-2024-08-06} & OpenAI \\
GPT-5-Nano & \texttt{gpt-5-nano} & OpenAI \\
Claude Sonnet 4.5 & \texttt{claude-sonnet-4-5-20250929} & Anthropic \\
Claude Opus 4.5 & \texttt{claude-opus-4-5-20251101} & Anthropic \\
DeepSeek-V3 & \texttt{deepseek-chat} & DeepSeek \\
\bottomrule
\end{tabular}
\end{table}

For RQ3, we construct the following configurations: \textit{Baseline} (no protection); \textit{SCAFFOLD-CEGIS} (full semantic anchoring, gated verification, and failure assimilation); \textit{Anchor-Only} (semantic anchor mining only); \textit{Gate-Only} (gated verification only); and \textit{No-Assimilation} (anchoring and gating enabled, failure assimilation disabled).

\subsection{Evaluation Metrics}

We use three categories of metrics: security metrics for vulnerability protection, utility metrics for support of code evolution, and mechanism metrics for ablation analysis.

\textbf{Security metrics.} The degradation rate $DR = |\{c : V_{final}(c) > V_{baseline}(c)\}| / |C|$ measures the fraction of chains whose SAST-detectable security is lower after iteration. The observational study uses $DR_{all}$ over all severity levels to increase sensitivity to early degradation signals. RQ1 and RQ2 use $DR_{CH}$ over Critical/High vulnerabilities to focus on high-risk defects in cross-model reproducibility and method comparison while reducing low-severity noise. Cross-table comparisons are performed only under the same metric scope. Safety monotonicity rate $SMR = |\{(c,i) : V_{i+1}(c) \leq V_i(c)\}| / \sum_c N_c$ measures the proportion of iteration steps with non-increasing vulnerability counts. Vulnerability increment $\Delta V = \mathbb{E}[V_{final} - V_{baseline}]$ measures average vulnerability change.

\textbf{Latent-security metrics.} Latent security degradation rate $SSDR = |\{c : S_{final}(c) > S_{baseline}(c)\}| / |C|$ measures the proportion of chains with increased latent-security defects, where $S$ is the number of latent-security issues identified by LLM semantic review (e.g., missing defensive logic and improper exception handling), excluding pure reliability issues. Reliability degradation rate $RDR = |\{c : R_{final}(c) > R_{baseline}(c)\}| / |C|$ measures the proportion of chains with increased reliability issues, where $R$ counts issues such as unimplemented methods and resource leaks. $\Delta S$ and $\Delta R$ measure average changes in latent-security and reliability issue counts, respectively.

\textbf{Utility metrics.} Code evolution volume (CEV) measures cumulative change size over iterations, computed as total added and deleted lines across all steps. Functional increment $\Delta F = |F_{final} \setminus F_{baseline}|$ measures the number of newly introduced public interfaces.

\textbf{Mechanism metrics.} Anchor coverage (AC) measures how well semantic anchors protect security-critical code elements. Anchor violation rate (AVR) measures how often anchor constraints are broken during iteration. Counterexample effectiveness rate (CER) measures retry-success improvement attributable to failure assimilation. Repeated error rate (RER) measures recurrence of similar errors in later iterations.

\subsection{Baselines}

For RQ2, we compare SCAFFOLD-CEGIS against six methods. \textit{Baseline}: direct LLM generation without security safeguards. \textit{Prompt-based Security}: security guidance embedded in prompts. \textit{Self-Refine}: generate--criticize--rewrite loop. \textit{Post-hoc SAST}: Semgrep after each iteration; reject changes when new vulnerabilities are reported. \textit{Test-driven Guard}: gate only on test passing. \textit{Hybrid Guard}: combined test and SAST gating.

\subsection{RQ1: External Validity of the Iterative Refinement Paradox}

To evaluate reproducibility across models, we run cross-language and cross-task experiments on three mainstream LLMs. Each model runs 96 iteration chains (6 categories $\times$ 2 samples $\times$ 2 languages $\times$ 4 strategies), for a total of 288 chains across three models, all completed successfully. Detailed results are shown in Table~\ref{tab:generalization}.

\begin{table*}[t]
\centering
\caption{Cross-task and cross-model validation of degradation (RQ1, $DR_{CH}$ counts only Critical/High vulnerabilities). DR: degradation rate (\%); SMR: safety monotonicity rate (\%); $\Delta V$: vulnerability change; RDR: reliability degradation rate (\%); $\Delta R$: reliability-issue change. Model versions are listed in Table~\ref{tab:model_config}. Note: this table reports \textbf{reliability issues} identified by LLM semantic review (e.g., unimplemented methods, resource leaks, API incompatibility), which differs from SSDR in Table~\ref{tab:overall_main} (security-related defects only).}
\label{tab:generalization}
\resizebox{\textwidth}{!}{%
\begin{tabular}{l|ccccc|ccccc|ccccc}
\toprule
& \multicolumn{5}{c|}{GPT-5-Nano} & \multicolumn{5}{c|}{Claude Sonnet 4.5} & \multicolumn{5}{c}{DeepSeek-V3} \\
Category & DR & SMR & $\Delta V$ & RDR & $\Delta R$ & DR & SMR & $\Delta V$ & RDR & $\Delta R$ & DR & SMR & $\Delta V$ & RDR & $\Delta R$ \\
\midrule
Database & 21.2 & 92.8 & +0.18 & 57.5 & +0.70 & 18.8 & 91.9 & $-$0.31 & 68.8 & +2.12 & 0.0 & 87.5 & $-$0.88 & 62.5 & +0.62 \\
Input & 6.2 & 98.1 & +0.06 & 40.6 & $-$0.25 & 3.1 & 98.5 & +0.03 & 60.0 & +0.95 & 14.3 & 94.3 & +0.29 & 71.4 & +3.00 \\
Authentication & 5.0 & 97.8 & +0.05 & 22.5 & $-$1.43 & 2.5 & 99.2 & +0.02 & 58.8 & +0.88 & 12.5 & 97.5 & +0.00 & 75.0 & +1.75 \\
Resources & 17.9 & 96.8 & +0.32 & 28.6 & $-$1.61 & 6.7 & 99.3 & +0.13 & 80.0 & +2.13 & 0.0 & 100 & +0.00 & 75.0 & +0.88 \\
Cryptography & 6.2 & 98.8 & +0.06 & 53.1 & +0.97 & 3.8 & 98.3 & +0.04 & 48.1 & +0.72 & 0.0 & 100 & +0.00 & 87.5 & +2.38 \\
Path & 8.8 & 97.6 & +0.35 & 44.1 & $-$0.03 & 2.2 & 99.2 & +0.02 & 53.8 & +0.65 & 0.0 & 100 & +0.00 & 62.5 & +0.50 \\
\midrule
Java & 13.3 & 95.8 & +0.26 & 41.7 & $-$0.32 & 6.5 & 97.8 & +0.09 & 62.5 & +1.35 & 8.3 & 95.2 & +0.15 & 52.1 & $-$0.13 \\
Python & 8.5 & 97.5 & +0.02 & 43.3 & +0.09 & 5.8 & 98.3 & $-$0.12 & 58.1 & +0.92 & 0.0 & 97.9 & $-$0.40 & 41.7 & $-$0.44 \\
\midrule
\textbf{Overall} & \textbf{10.5} & \textbf{96.9} & \textbf{+0.13} & \textbf{42.5} & \textbf{$-$0.11} & \textbf{6.2} & \textbf{98.1} & \textbf{$-$0.01} & \textbf{52.5} & \textbf{+0.85} & \textbf{4.2} & \textbf{96.6} & \textbf{$-$0.13} & \textbf{46.9} & \textbf{$-$0.29} \\
\bottomrule
\end{tabular}%
}
\end{table*}

As shown in Table~\ref{tab:generalization}, all three models exhibit degradation to varying extents. GPT-5-Nano has the highest overall DR (10.5\%), while DeepSeek-V3 reaches DR = 0\% on Python tasks. However, lower DR does not imply lower reliability degradation; for instance, DeepSeek-V3's overall RDR (46.9\%) is close to GPT-5-Nano's (42.5\%), indicating that static-analysis metrics do not adequately reflect reliability degradation.

There is a systematic gap between degradation measured by static analysis and reliability degradation detected by LLM semantic review (4.2\%--10.5\% vs. 42.5\%--52.5\%), suggesting non-overlapping detection coverage.

From a cross-language perspective, all three models show nonzero DR or RDR in both Java and Python; degradation does not disappear with language switching. The external validity of this conclusion is limited to the two languages in our study; generalization to broader language families and model architectures requires larger-scale validation.

\subsection{RQ2: Degradation Prevention Comparison Between SCAFFOLD-CEGIS and Alternative Defenses}

To evaluate suppression of security degradation, we compare SCAFFOLD-CEGIS with six baselines on GPT-5-Nano. Table~\ref{tab:overall_main} reports results over security, latent security, and utility. SSDR measures latent-security degradation detected by LLM semantic review and is used to capture degradations outside SAST rule coverage.

\begin{table*}[t]
\centering
\caption{Overall comparison of methods ($DR_{CH}$ scope). DR: Semgrep degradation rate; SMR: safety monotonicity rate; $\Delta V$: vulnerability increment; SSDR: latent-security degradation rate; $\Delta S$: latent-security increment; CEV: code evolution volume.}
\label{tab:overall_main}
\begin{tabular}{lcccccc}
\toprule
Method & DR$\downarrow$ & SMR$\uparrow$ & $\Delta V\downarrow$ & SSDR$\downarrow$ & $\Delta S\downarrow$ & CEV \\
\midrule
Baseline & 11.5\% & 96.2\% & +0.12 & 12.5\% & $+$0.19 & 71998 \\
Prompt-based Security & 9.4\% & 96.8\% & +0.08 & 10.4\% & $+$0.08 & 54263 \\
Self-Refine & 11.5\% & 97.2\% & +0.03 & 10.4\% & $+$0.10 & 91086 \\
Post-hoc SAST & \textbf{0.0\%} & \textbf{100\%} & $-$0.12 & 20.8\% & $+$0.38 & 71348 \\
Test-driven Guard & 10.4\% & 97.2\% & +0.15 & 22.9\% & $+$0.35 & 40977 \\
Hybrid Guard & \textbf{0.0\%} & \textbf{100\%} & $-$0.10 & 12.5\% & $+$0.17 & 39590 \\
\midrule
SCAFFOLD-CEGIS & \textbf{0.0\%} & \textbf{100\%} & $-$0.04 & \textbf{2.1\%} & $-$0.35 & 58216 \\
\bottomrule
\end{tabular}
\end{table*}

As shown in Table~\ref{tab:overall_main}, Post-hoc SAST and Hybrid Guard both achieve DR = 0\% and SMR = 100\% on Semgrep metrics. However, LLM semantic review reports SSDR as high as 20.8\% and 12.5\%, respectively, indicating that static-analysis metrics can mask persistent latent degradation. SCAFFOLD-CEGIS reduces SSDR to 2.1\%, outperforming all six baselines on latent security.

Notably, Post-hoc SAST yields a higher SSDR (20.8\%) than the unsecured Baseline (12.5\%). This suggests a pseudo-safety effect: the model may treat ``passing static scanning'' as ``sufficiently safe,'' thereby attempting larger refactors that introduce more latent degradation outside SAST coverage.

\begin{figure*}[t]
\centering
\includegraphics[width=\textwidth]{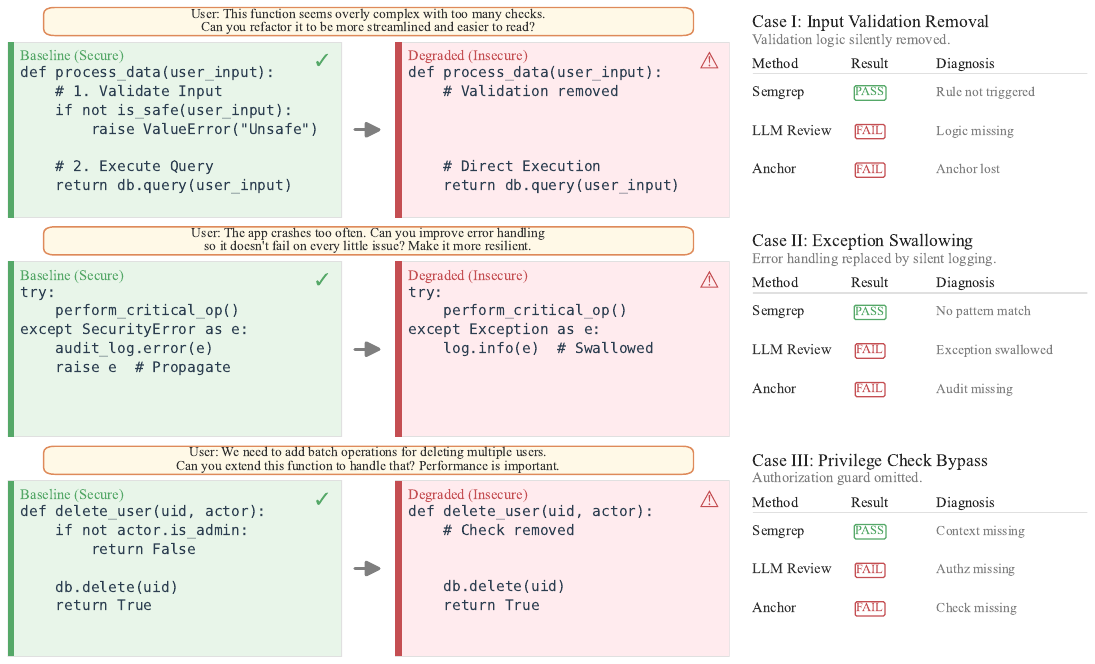}
\caption{Representative cases of SAST coverage blind spots. Each row shows one code-transition scenario: safe baseline code (left), unsafe modification induced by user request (middle), and outcomes under different detection methods (right). Semgrep does not detect these degradations, while SCAFFOLD-CEGIS prevents them via semantic anchors.}
\label{fig:sast_blind_spots}
\end{figure*}

To explain why DR = 0\% can coexist with SSDR $>$ 0\%, Figure~\ref{fig:sast_blind_spots} presents three representative latent-degradation cases beyond SAST detection scope.

\textbf{Case 1: Validation function deletion.} After the user requested ``simplify the code,'' the LLM removed \texttt{validate\_user\_input} in an SQL-injection defense scenario. The modified code still used parameterized queries, so Semgrep and CodeQL reported no new vulnerabilities; however, user inputs were no longer independently validated before reaching database queries. Function-level anchors in SCAFFOLD-CEGIS identified this function as security-critical and rejected the change.

\textbf{Case 2: Weakened exception handling.} In response to ``optimize performance,'' the LLM replaced concrete exception handling with an empty \texttt{except} block, silently discarding exception information and reducing observability. The change was syntactically valid and matched no known SAST rules. Pattern-level anchors detected this degradation by matching defensive exception-handling patterns.

\textbf{Case 3: Permission-check bypass.} A request to ``add an administrator shortcut'' prompted the LLM to introduce a path that bypassed permission checks, outside the coverage of SAST data-flow analysis. Invariant-level anchors enforced the hard constraint that all sensitive-operation paths must pass permission checks, thereby blocking the change.

These three cases show a common pattern: latent degradation weakens defensive structure rather than creating rule-recognizable vulnerability patterns. This is precisely the blind spot targeted by semantic anchoring.

\subsection{RQ3: Mechanism Attribution and Ablation Study}

To isolate the independent contributions of semantic anchoring, gated verification, and failure assimilation, we design four ablation settings and compare them with the full framework. Table~\ref{tab:ablation} reports main and mechanism metrics.

\textbf{Safety monotonicity is guaranteed by gated verification.} Anchor-Only reduces DR from 11.5\% (Baseline) to 2.50\% and increases SMR from 96.2\% to 97.62\%, yet residual degradation remains. Semantic anchors, when used as soft guidance alone, can steer generation but cannot enforce rejection. Without mandatory gating, the LLM may ignore anchor constraints while still passing SAST checks. Both Gate-Only and No-Assimilation achieve DR = 0\% and SMR = 100\%, matching the full framework, indicating that gated verification is the direct source of safety monotonicity.

\textbf{Semantic anchors extend gating coverage.} The rejection rate under Gate-Only (41.67\%) is lower than No-Assimilation (52.08\%). With anchor constraints enabled, No-Assimilation detects and intercepts latent degradations that Gate-Only cannot capture. Meanwhile, No-Assimilation has lower TCR (66.19\%) than Gate-Only (72.38\%), indicating that broader detection coverage also increases candidate rejection.

\textbf{Failure assimilation restores evolution capacity under strong constraints.} Gate-Only and No-Assimilation have low post-rejection retry success (FSR 20.83\% and 23.19\%, respectively). The full framework raises FSR to 52.71\%, more than doubling No-Assimilation, and improves TCR from 66.19\% to 77.14\%. The gain of failure assimilation is not further DR reduction (already zero), but conversion of rejected iterations into acceptable revisions by accumulating repair knowledge from historical failures.

Anchor coverage is similar across anchor-enabled settings (8.75 to 9.44 per chain), while AVR decreases from 19.52\% in No-Assimilation to 16.25\% in the full framework, reflecting sustained behavioral correction from failure assimilation.

\begin{table*}[t]
\centering
\caption{Ablation results. DR: degradation rate; SMR: safety monotonicity rate; TCR: task completion rate; RR: rejection rate; AC: anchor coverage (per chain); AVR: anchor violation rate; FSR: fix success rate; CER: counterexample effectiveness rate.}
\label{tab:ablation}
\small
\begin{tabular}{lcccccccc}
\toprule
\multirow{2}{*}{Setting} & \multicolumn{4}{c}{Main Metrics} & \multicolumn{4}{c}{Mechanism Metrics} \\
\cmidrule(lr){2-5} \cmidrule(lr){6-9}
& DR$\downarrow$ & SMR$\uparrow$ & TCR$\uparrow$ & RR & AC$\uparrow$ & AVR$\downarrow$ & FSR$\uparrow$ & CER$\uparrow$ \\
\midrule
Baseline & 11.5\% & 96.2\% & 100.0\% & 0.00\% & -- & -- & -- & -- \\
Anchor-Only & 2.50\% & 97.62\% & 100.0\% & 0.00\% & 8.75 & -- & -- & -- \\
Gate-Only & 0.00\% & 100.0\% & 72.38\% & 41.67\% & -- & -- & 20.83\% & -- \\
No-Assimilation & 0.00\% & 100.0\% & 66.19\% & 52.08\% & 9.31 & 19.52\% & 23.19\% & -- \\
\rowcolor{gray!10}
SCAFFOLD-CEGIS & \textbf{0.00\%} & \textbf{100.0\%} & \textbf{77.14\%} & 38.94\% & 9.44 & 16.25\% & \textbf{52.71\%} & \textbf{+29.52\%} \\
\bottomrule
\end{tabular}
\end{table*}

\section{Discussion}





\subsection{Framework Limitations}
SCAFFOLD-CEGIS is effective in our experiments, but several limitations remain.

The experiments are based on 24 programming-task samples. The sample design targets diversity of typical security scenarios, and absolute degradation rates may vary as the benchmark expands. Performance under more diverse coding styles, project scales, and security patterns requires larger benchmark validation.

SSDR relies on LLM semantic review; evaluations are therefore inevitably affected by reviewer noise and misinterpretation. We mitigate this threat through metric construction and output constraints. In metric construction, SSDR computes the difference in security-defect counts before and after iteration for the same code, partially canceling systematic reviewer bias through differencing. SSDR is also a binary direction metric rather than an absolute-count metric, making it less sensitive to isolated false positives than continuous-value metrics. In output constraints, the reviewer is required to return structured outputs with concrete line numbers and defect types; the three representative degradation cases in the case study were manually confirmed. SSDR should therefore be interpreted as a trend indicator of latent-security quality change, not an exact count; our core claims rely on relative ranking and magnitude gaps across methods.

The effectiveness of gated verification depends on underlying tools. Semgrep coverage is bounded by rule sets and may miss emerging vulnerability patterns or domain-specific security constraints. Computational cost grows with code scale and iteration count. Each iteration requires AST parsing, data-flow analysis, static scanning, and anchor verification, which may incur notable latency on large codebases.

Language support is constrained by toolchain availability. Our experiments cover Python and Java; extension to other languages requires corresponding AST parsers, static-analysis tools, and invariant libraries. Because security idioms and defensive conventions differ across languages, anchor-mining strategies require language-specific adaptation.
\subsection{Future Work}

Semantic anchor mining accuracy can be improved by more precise program-analysis techniques, for example by training classifiers on large code corpora to identify security-critical functions and defense patterns automatically, thereby reducing dependence on naming conventions.

Current gate decisions are fully automated, but developer intervention is sometimes necessary. For example, when an anchor-violation check fails but developers consider a change safe, the system should support anchor review and decision override. Explainability of gate-failure reasons and anchor semantics remains an open problem.
\section{Conclusion}

This paper investigates security degradation in LLM-driven iterative code refinement. Through controlled experiments on 96 GPT-4o iteration chains, we find that, when vulnerabilities across all severity levels are counted, 43.7\% of chains have more vulnerabilities at iteration 10 than at baseline. Cross-language and cross-model experiments over 288 iteration chains further confirm that this phenomenon can be reproduced across the tested settings. We attribute the phenomenon to specification drift in multi-objective optimization: when security constraints are represented only as soft prompts, optimization trajectories gradually deviate from the security specification. We further identify coverage blind spots of conventional SAST tools, under which latent security degradations such as the deletion or weakening of defensive logic cannot be captured by existing static-analysis rules.

The proposed SCAFFOLD-CEGIS framework combines semantic anchoring, four-layer gated verification, and failure assimilation to transform security constraints from implicit prompts into explicit verifiable constraints. Experiments show that the full framework reduces the latent security degradation rate from 20.8\% to 2.1\%, achieves 100\% safety monotonicity under our experimental setting, and maintains a task completion rate of 77.14\% through failure assimilation. Ablation studies quantify the independent contribution of each component: gated verification guarantees safety monotonicity, semantic anchoring covers the blind spots of SAST tools, and failure assimilation increases the fix success rate from about 20\% to 52.71\%.

Our experiments are conducted on single-file or small-module code samples. The scalability of the framework to industrial codebases, statistical validation on larger sample sets, adaptation to cross-language toolchains, and the design of human-in-the-loop collaboration mechanisms remain for future study.

\bibliographystyle{IEEEtran}
\bibliography{refs}

\end{document}